\documentclass{aa}
\usepackage{color}
\usepackage{xcolor}
\usepackage{listings}
\usepackage[frozencache]{minted}
\usepackage{amsmath}
\usepackage{natbib}
\usepackage{multirow}
\usepackage{booktabs}
\newcommand{\vaex}{\texttt{vaex}\xspace}
\newcommand{\vaexui}{\texttt{vaex-ui}\xspace}
\newcommand{\Vaex}{\texttt{Vaex}\xspace}
\newcommand{\Python}{\texttt{Python}\xspace}
\newcommand{\Pandas}{\texttt{Pandas}\xspace}
\newcommand{\Topcat}{\texttt{TOPCAT}\xspace}
\newcommand{\hdf}{\texttt{hdf5}\xspace}
\newcommand{\fits}{\texttt{FITS}\xspace}
\newcommand{\colfits}{\texttt{col-fits}\xspace}
\newcommand{\matplotlib}{\texttt{matplotlib}\xspace}
\newcommand{\numpy}{\texttt{numpy}\xspace}
\newcommand{\Clang}{\texttt{C}\xspace}
\newcommand{\Cpplang}{\texttt{C++}\xspace}
\newcommand{\Lz}{L$_z$\xspace}
\newcommand{\Energy}{E\xspace}
\newcommand{\Ltot}{L\xspace}

\begin{document}

\title{Vaex: Big Data exploration in the era of Gaia}

\author{Maarten A. Breddels
  \and
  Jovan Veljanoski
}
\institute{Kapteyn Astronomical Institute, University of Groningen, P.O. Box 800, 9700 AV Groningen, The Netherlands}
\authorrunning{M.A. Breddels \& J. Veljanoski}
\abstract{We present a new Python library called \vaex{}, to handle extremely
large tabular datasets, such as astronomical catalogues like the Gaia catalogue, N-body simulations or
any other regular datasets which can be structured in rows and columns. Fast
computations of statistics on regular N-dimensional grids allows analysis and
visualization in the order of a billion rows per second. We use streaming
algorithms, memory mapped files and a zero memory copy policy to allow
exploration of datasets larger than memory, e.g. out-of-core algorithms.
\Vaex{} allows arbitrary (mathematical) transformations using normal Python
expressions and (a subset of) \numpy functions which are lazily evaluated and
computed when needed in small chunks, which avoids wasting of RAM. Boolean
expressions (which are also lazily evaluated) can be used to explore subsets of
the data, which we call selections. \Vaex{} uses a similar DataFrame API as
Pandas, a very popular library, which helps migration from Pandas. Visualization
is one of the key points of \vaex{}, and is done using binned statistics in 1d
(e.g. histogram), in 2d (e.g. 2d histograms with colormapping) and 3d (using
volume rendering). \Vaex{} is split in in several packages: \texttt{vaex-core}
for the computational part, \texttt{vaex-viz} for visualization mostly based on
matplotlib, \texttt{vaex-jupyter} for visualization in the Jupyter notebook/lab
based in IPyWidgets, \texttt{vaex-server} for the (optional) client-server
communication, \texttt{vaex-ui} for the Qt based interface,  \texttt{vaex-hdf5}
for hdf5 based memory mapped storage, \texttt{vaex-astro} for astronomy related
selections, transformations and memory mapped (column based) fits storage. \Vaex
is open source and available under MIT license on github, documentation and
other information can be found on the main website: \url{https://vaex.io} or
\url{https://github.com/maartenbreddels/vaex}}

\maketitle

\section{Introduction}
\label{sec:intro}

Visualization plays an important role in astronomy, and is often used to find
and display trends in data in the form of two dimensional scatter plots.
The Herzsprung-Russel diagram for example, is one of the most famous scatter
plots, which shows the relationship between the temperature and the luminosity
of stars. Before the era of computers, these plots were drawn by hand, while now
it is customary to use a software package or a library to produce them.

While two dimensional scatter plots may reveal trends or structure in a dataset
of relatively small size, they become illegible when the number of samples
exceeds $\sim 10^6$: the symbols overlap and ultimately fill up the plot in a
uniform colour, obscuring any information such a figure may contain. When a
dataset contains more than $10^6$ samples, it is more meaningful to visualize
the local density in a two dimensional plot. The density can be determined using
a kernel density estimator (KDE) or by a binning technique, equivalent to
constructing a histogram in one dimension. The value of the local density can
then be translated to a colour using a colourmap, which makes for an informative
visualization.

\begin{figure*}[t]
\begin{center}
\begin{tabular}{cc}
\includegraphics[scale=0.3]{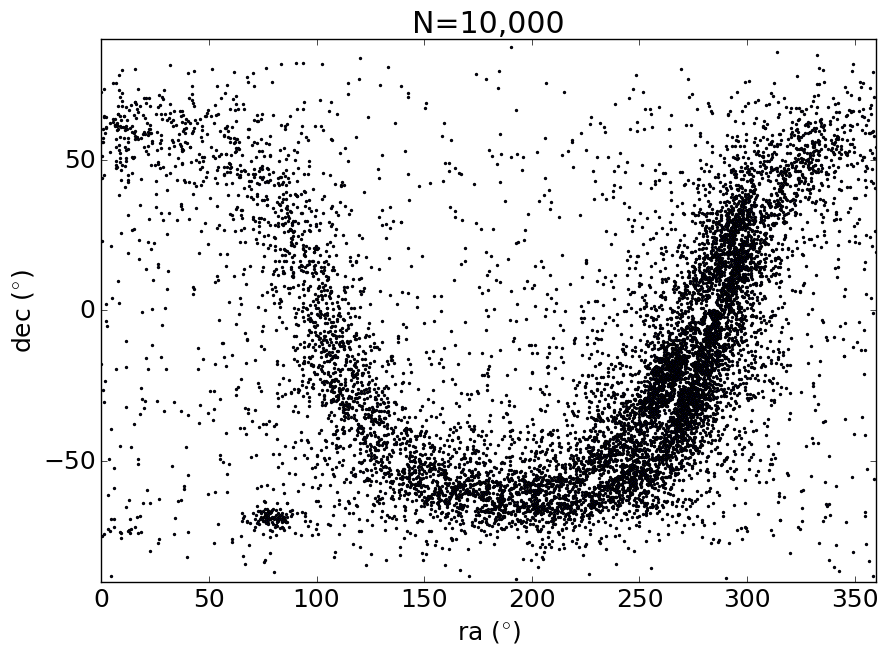} &
\includegraphics[scale=0.3]{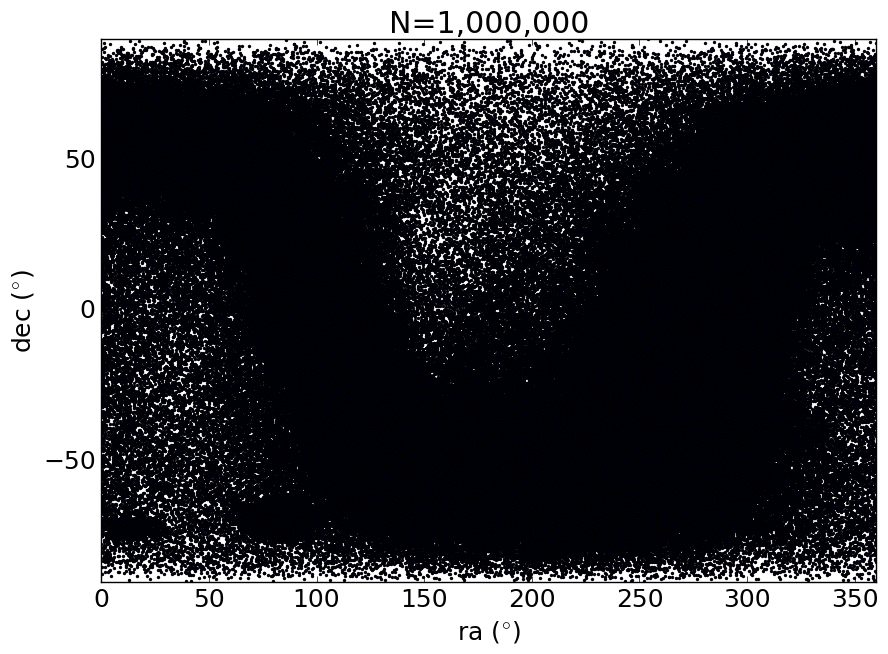} \\
\multicolumn{2}{c}{\includegraphics[scale=0.3]{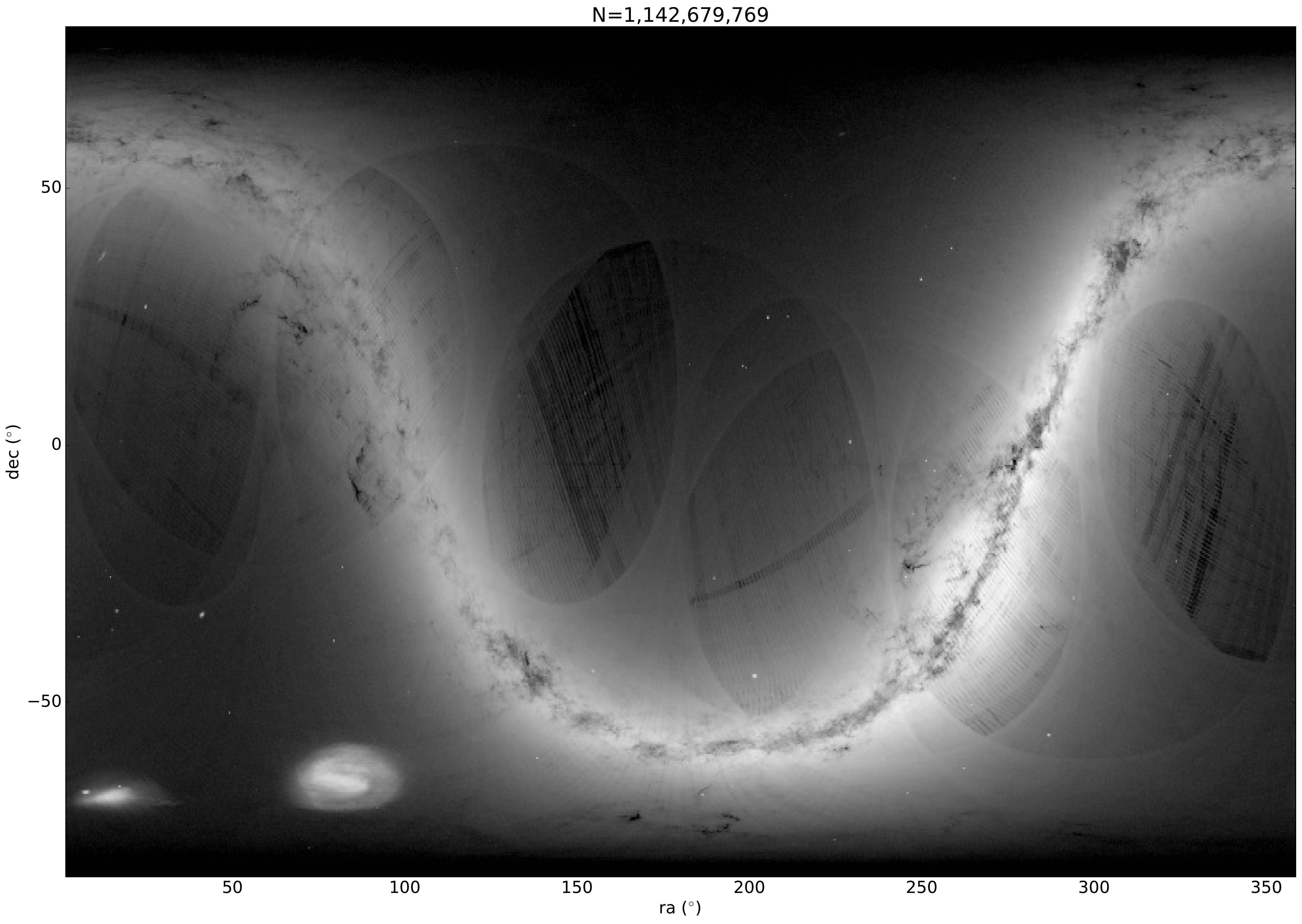}}
\end{tabular}
\caption{Comparison between scatter and density plots when trying to visualize
the sky as observed by \emph{Gaia}~DR1. \textbf{Top left:} A scatter plot
showing $10\,000$ sources in Equatorial coordinates, which does reveal some
structure in the disk. \textbf{Top right:} Idem, with $1\,000\,000$ sources,
hiding almost any structure present in the data. \textbf{Bottom:} A density plot
with $1\,142\,679\,769$ sources, where the logarithm of the density is
colour-mapped (black is low density and white is high density). This reveals
much more features in the data, such as structure in the disk and artefacts
related to the scanning nature of the \emph{Gaia} satellite.}
\label{fig:scatter_vs_density}
\end{center}
\end{figure*}

To illustrate this concept, in Figure~\ref{fig:scatter_vs_density} we show the
positions in equatorial coordinates of the stars in the \emph{Gaia}~DR1
catalogue \citep{GaiaDR1cat}, which contains over 1~billion entries in total.
On the top left panel we show a scatter plot containing only $10^4$ randomly
chosen stars. This plot shows some structure, the Galactic disk is clearly
seen, and one can also see the Large Magellanic Cloud as an over-density of
points at (ra, dec) $\approx (80,-70)$. These structures are largely smeared out
and nearly unnoticeable on the right panel where we show a scatter plot with
$1\,000\,000$ stars. On the other hand, we get significantly more information if
we visualize the data with a density plot. The bottom panel in
Figure~\ref{fig:scatter_vs_density} shows a density plot of the entire
\emph{Gaia}~DR1 catalogue, where one can see in great detail the structure of
the disk, the Magellanic Clouds, patterns related to how the satellite scans
the sky, and even some dwarf galaxies and globular clusters. All these details
are lost when we represent the data with a scatter plot. However, a
visualization library cannot stand on its own, and needs additional support for
efficient transformation, filtering and storing of the data, as well as
efficient algorithms to calculate statistics that form the basis of the
visualization.

In this paper we present a new \Python{} library called \vaex, which is able to
handle extremely large tabular datasets such as astronomical catalogues, N-body
simulations or any other regular datasets which can be structured in rows and
columns. Fast computations of statistics on regular N-dimensional grids allows
analysis and visualizations in the order of a billion rows per second. We use
streaming algorithms, memory mapped files and a zero memory copy policy to allow
exploration of datasets larger than the Read Access Memory (RAM) of a computer
would normally allow, e.g. out-of-core algorithms. \Vaex{} allows arbitrary
mathematical transformations using normal \Python{} expressions and \numpy{}
functions which are lazily evaluated, meaning that they are only computed when
needed, and this is done in small chunks which optimizes the RAM usage.
Boolean expressions, which are also lazily evaluated, can be used to explore
subsets of the data, which we call selections. \Vaex{} uses a similar DataFrame
API as \Pandas{} \citep{PandasMckinney-proc-scipy-2010}, a very popular
\Python{} library, which lessens the learning curve and makes its usage more
intuitive. Visualization is one of the focus points of \vaex{}, and is done
using binned statistics in one dimension (histogram), in two dimensions (2d
histograms with colour-mapping) and in three dimensions (using volume
rendering). \Vaex{} is split in several packages: \texttt{vaex-core} for the
computational part, \texttt{vaex-viz} for visualization mostly based on
matplotlib, \texttt{vaex-server} for the optional client-server communication,
\texttt{vaex-ui} for the Qt based interface, \texttt{vaex-jupyter} for
interactive visualization in the Jupyter notebook/lab based on \texttt{IPyWidgets},
\texttt{vaex-hdf5} for \hdf based memory mapped storage and \texttt{vaex-astro}
for astronomy related selections, transformations and \texttt{(col)fits}
storage.

Other similar libraries or programs exist, but do not match the performance or
capabilities of \vaex. \Topcat \citep{Topcat2005ASPC}, a common tool in
astronomy, has support for density maps, but in general is focussed on working
on a per row basis, and does not handle $10^9$ objects efficiently. The
\Pandas{} library can be used for similar purposes, but its focus is on in
memory data structures. The
\texttt{datashader}\footnote{\url{https://github.com/bokeh/datashader}} library
can handle large volumes of data, focuses mainly on visualization in two
dimensions, and lacks tools for exploration of the data.
\texttt{Dask}\footnote{\url{https://dask.pydata.org/}} and especially its
\texttt{DataFrame} library is a good alternative for the computational part of
\vaex{} but it is accompanied with a rather steep learning curve.

This paper is structured as follows. In Section~\ref{sec:main} we begin by
laying out the main ideas that form \vaex, which we support with the relevant
calculations. In Section~\ref{sec:library}, we first present the basis of the
library (\texttt{vaex-core}), and discuss all other packages that are
subsequently built on top of it, such as \texttt{vaex-astro} for the astronomy
related package, and \texttt{vaex-ui} for the Qt based user interface. We
summarize our work in Section~\ref{sec:conclusions}.

Note that this paper does not document all the features and options of the
the \vaex library. It lays out the principle ideas and motivation for creating
the software, and presents the main capabilities of \vaex. The full
documentation can be found at: \url{https://vaex.io}.  \Vaex is open source and available under the MIT licence on  github at: \url{https://github.com/maartenbreddels/vaex}.

\section{Main ideas}
\label{sec:main}

In this section, we lay out the main ideas and the motivation for developing
\vaex. We start by discussing the possibilities and limitations when dealing
with large tabular datasets. We then present some calculations to show that it
is indeed theoretically possible to process 1 billion samples per second,
and reflect back on that with an implementation.

\subsection{Constraints and possibilities}

In the Introduction we clearly showed how a scatter plot displaying $\sim 10^9$
samples is usually not meaningful due to over-plotting of the symbols. In
addition, when one want to process $10^9$ samples in one second on a Intel(R)
Core(TM) i7-4770S CPU $3.1$ GHz machine with four cores, only $12.4$ CPU cycles
are available per sample. That does not leave room for plotting even one glyph
per object, as only a few CPU instructions are available. Furthermore,
considering numerical data for two columns of the double precision floating
point type, the memory usage is 16~GB $(10^9\times 2 \times 8~\text{bytes} =
16\times 10^9~\text{bytes} \approx 15~\text{GiB})$, which is quite large
compared to a maximum bandwidth of $~25.6 \text{GB/s}$ for the same CPU.

Therefore, for the \vaex library we only consider streaming or out-of-core
algorithms which need one or a few passes over the data, and require few
instructions per sample. The computation of statistics such as the mean or
higher moments of the data are examples of such algorithms. The computation of
a histogram on a regular grid can also be done with only few instructions,
enabling us to efficiently visualize large amounts of data in one dimension by
means of a histogram, in two dimensions via a density plot, and in three
dimensions by the means of volume or isosurface rendering.

Preprocessing the data can lead to additional increase in performance. Given
that users often perform various transformations on the data while they are
exploring it, such as taking the log of a quantity or the difference between two
quantities, we do not consider any preprocessing.

\newcommand{\processAmount}{0.6 billion\xspace}
\newcommand{\processTime}{0.55 seconds\xspace}
\newcommand{\processSpeed}{1.1 billion objects/s\xspace}

\subsection{Real performance}

We implemented a simple binning algorithm in
\Clang with a Python binding, finding that we can create a $256\times256$ two
dimensional histogram from a dataset with \processAmount samples in
\processTime, processing \processSpeed, which we consider acceptable for
interactive visualization. This code uses multi-threading\footnote{Releasing
Python's Global Interpreter Lock when entering the \Clang part to actually make
use of the multi-threading.} to achieve this speed, while using $\sim 75-85\%$
(15-17~GB/s) of the maximum memory bandwidth\footnote{Although the theoretical
bandwidth is 25~GB/s, we measured it to be 20~GB/s using the bandwidth program
from \url{http://zsmith.co/bandwidth.html}}.

\subsection{N-dimensional statistics}

Apart from simply counting the number of samples in each bin, one can generalize
this idea to calculate other statistics per bin using extra columns. Instead of
simply summing up the number of samples that fall into each bin, one can use the
same algorithm to perform other computations on a particular column, effectively
calculating many statistics on a regular grid in N dimensions, where 0
dimensions implies a scalar. For example, let us consider a dataset that
features four columns {$x$, $y$, $v_x$, $v_y$}, where the first two represent the
position and the last two the corresponding velocity components of a particle
or a star. One can construct a two dimensional grid spanned by $x$ and $y$
displaying the mean $v_x$ by first summing up the $v_x$ values and then dividing
by the total number of samples that fall into each bin. The same exercise can be
repeated to calculate the mean velocity in $y$ direction. Higher order moment
can also be calculated, allowing one to compute and visualize vector and tensor
quantities in two and three dimensions. The types of statistics available in
\vaex are listed in Section~\ref{sec:statistics} and Table~\ref{tab:algo}.

\subsection{Implementation}

These ideas and algorithms, which are efficiently implemented, form the basis
of the \vaex library. \Vaex exposes them in a simple way, allowing users to
perform computations and scientific visualizations of their data with minimal
amount of code. The graphical user interface program, from now on referred to as
the program, uses the library to directly visualize datasets to users, and
allows for interactive exploration. By this we mean the user is not only able to
navigate (zoom and pan), but also to make interactive selections (visual
queries), which can be viewed in other windows that display a plot created by a
different combination of columns from those on which the selection was made on
(linked views). \Vaex also provides ranking of subspaces\footnote{We call a
combination of 1 or more columns (or expression using columns) a subspace},
by calculating their mutual information or correlation coefficient in order to
find out which subspaces contain more information.

\subsection{Using a part of the data}

In some cases, it may be useful to do computations a smaller random subset of
of data. This is beneficial for devices that do not have enough storage to keep
the whole dataset such as laptops, and will also require less computing power.
This is also useful for servers, as we will see in Section~\ref{sec:vaex-server}
in order to handle many requests per second. Instead of drawing a random subset
of rows the the full dataset, we store the dataset with the rows in a random
order, and than `draw' a random subset of rows (which will be the same
every time), by only processing the first N rows. To support this, the library
includes the option to covert and export a dataset with the rows in a random
order. Note that to shuffle more than $2^{32}\approx4.2\times10^9$ rows, a 64
bit random number generator is needed. For the moment, this is only supported
on the Linux operation system.

\section{\Vaex}
\label{sec:library}

The ideas of the previous section form the basis of the \vaex library. \Vaex is
a \Python package, consisting of pure \Python modules as well as a so called
extension module, written in \Clang, which contains the fast performing
algorithms, such as those for binning the data. The \vaex library can be
installed using \texttt{pip}, or (ana)conda\footnote{A popular Python
distribution: \url{https://www.continuum.io/downloads}.}. Its source code and
issue tracker are on-line at \url{https://github.com/maartenbreddels/vaex},
and the homepage is at \url{https://vaex.io}.

\Vaex is available as one (meta) package which will install all packages in the
\vaex family. However, if only a few functionalities are needed, only the
relevant packages can be installed. For instance in many cases only
\texttt{vaex-core}, \texttt{vaex-hdf5} and \texttt{vaex-viz} are needed. One can
thus avoid installing \texttt{vaex-ui} since it has (Py)Qt as a dependency,
which can be more difficult to install on some platforms.

\subsection{vaex-core}

The foundation of all \vaex packages is \texttt{vaex-core}. This contains the
most important part, the Dataset class, which wraps a series of columns
(\numpy arrays) in an API similar to Pandas' DataFrames, and gives access to all
the operations that can be performed on them, such as calculating statistics on
N-dimensional grids or the joining of two tables. On top of that, the Dataset
class does bookkeeping to track virtual columns, selections and filtering. Note
that in \vaex almost no operation makes copies of the data, since we expect the
full dataset to be larger than the RAM of typical computer.

\subsubsection{(Lazy) Expressions}

In practice, one rarely works only with the columns as they are stored in the
table. Within the \vaex framework, every statistic is based on a mathematical
expression, making it possible to not just plot the logarithm of a quantity
for example, but to plot and compute statistics using an arbitrary, often user
defined expression. For instance, there is no difference in usage when
calculating statistics on existing columns, for example the mean of $x$, or
any mathematical operation using existing columns, for example $x+y$, where
$x$ and $y$ are two columns of a \vaex dataset. The last expression will be
calculated on the fly using small chunks of the data in order to minimize memory
impact, and optimally make use of the CPU cache. Being able to calculate
statistics on an N-dimensional grid for arbitrary expressions is crucial for
exploring large datasets, such as the modern astronomical catalogues or outputs
of large-scale numerical simulations. For instance taking the logarithm of a
column is quite common, as well as calculating vector lengths
(e.g. $\sqrt{x^2+y^2+z^2}$). No pre-computations are needed, giving users the
complete freedom of what to plot or compute.

Contrary to the common \Pandas library, a statement like \mintinline{python}{a
= df.b + np.sin(df.c)}, where df is a \Pandas DataFrame containing the columns
\texttt{b} and \texttt{c}, would be directly computed and will results in
additional memory usage equal to that of the columns \texttt{b} or \texttt{c}.
In \vaex, the statement \mintinline{python}{a = ds.b + np.sin(ds.c)}, where
\texttt{ds} is a \vaex Dataset, results in an expression which only stores the
information of how the computation should be done. The expression will only be
calculated when the result of \texttt{a} is needed, which if often referred to
as a lazy evaluation. For convenience, a \vaex Dataset can also hold what we
refer to as virtual columns, which is a column that does not refer to a \numpy
array, but is an expression. This means that many columns can be added to a
dataset, without causing additional memory usage, and in many cases causing
hardly any performance penalty (when we are not CPU-bound). For instance, a
vector length can be added using \mintinline{python}{ds['r'] = np.sqrt(ds.x**2 +
ds.y**2 + ds.z**2)}, which can then be used in subsequent expressions. This
minimizes the amount of code that needs to be written and thus leads to less
mistakes.

\subsubsection{Just-in-time compilation}

Once the result of an expressions is needed, it is evaluated using the \numpy
library. For complex expressions this can result in the creation of many
temporary arrays, which may decrease performance. In these cases, the
computation speed of such complex expressions can be improved using
just-in-time (JIT) compilation by utilizing the \texttt{Pythran} \citep{Pythran}
or \texttt{Numba} \citep{Numba} libraries which optimize the code at runtime. Note that the
JIT compilation will not be done automatically, but needs to be manually applied
on an expression, e.g. \mintinline{python}{ds['a'] = (ds.x + ds.y + ds.z *
ds.w).jit()}.

\begin{table*}
\centering
\caption{A table listing the functions which can be computed on N dimensional
grids and exploit the fast binning algorithm, which are readily available in
\vaex. All statistics can be computed for the full dataset, a subset using selections or multiple selections at the same time. For all calculations, missing values or NaN's are ignored.}
\label{tab:algo}
\begin{tabular}{ll}
\hline
Statistic & Description \\
\hline
count & Counts the number of rows, or non-missing values of an expression. \\
sum & Sum of non-missing values of an expression. \\
mean & The sample mean of an expression. \\
var & The sample variance of an expression, using a non-stable algorithm. \\
std & The sample standard deviation of an expression using a non-stable algorithm. \\
min & The minimum value of an expression. \\
max & The maximum value of an expression. \\
minmax & The minimum and maximum value of an expression (faster than min and max seperately). \\
covar & The sample covariance between two expressions. \\
correlation & The sample correlation coefficient between two expressions, i.e. $\text{cov}[{x,y}] / \left(\sqrt{\text{var}[x]\text{var}[y]} \right)$.\\
cov & The full covariance matrix for a list of expressions. \\
percentile\_approx & \multirow{3}{*}{\parbox{12cm}{Estimates the percentile of an expression. Since the true value requires sorting of values, we implement an approximation by interpolation over a cumulative histogram.}} \\
\\
\\
median\_approx & Approximation of the median, based on the percentile statistic. \\
mode & Estimates the mode of an expression by calculating the peak of its histogram. \\
mutual\_information & Calculates the mutual information for two or more expression, see Section~\ref{sec:mi} for details. \\
nearest & Finds the nearest row to a particular point for given a metric. \\
\hline
\end{tabular}
\end{table*}

\subsubsection{Selections/Filtering}

In many cases one wants to visualize or perform computations on a specific
subset of the data. This is implemented by doing so called `selections' on the
dataset, which are one or more boolean expressions combined with boolean
operators. Such selections can be defined in two ways, via boolean expressions,
or via a geometrical (lasso) selection. The boolean expressions have the same
freedom as expressions applied on the dataset when computing statistics, and can
be a combination of any valid logical expressions supported by \Python using one
or more (virtual) columns. Examples of such booleans expressions are
\mintinline{python}{np.sqrt(ds.x**2 + ds.y**2 + ds.z**2) < 10} or
\mintinline{python}{(ds.FeH > -2) & (ds.FeH < -1)},  where the ampersand means
logical ``and''. Although the geometrical lasso selection can be implemented via
boolean expressions, it is implemented separately for performance reasons. The
lasso selection can be used a in graphical user interfaces to
select regions with a mouse or other pointing device, or to efficiently
select complex regions in two dimensions such as geographical regions. A dataset can have
multiple selections, and statistics or visualizations can be computed for one or
more selections at the same time (e.g. \mintinline{python}{ds.mean(ds.x,
selection=[ds.x<0, ds.y<0])}) in a single pass over the data. When selections are created using `select' method
\mintinline{python}{ds.select(selection=ds.x < 0, name='negative')}, they can
be named by passing a string to the name argument, and the result of the
selection, which is a boolean array, will be cached in memory leading to a
performance increase. If no name is given, it will assume the name 'default'.
Thus all selection arguments in \vaex can take a boolean expression as argument,
a name (referring to a selection made previously with
\mintinline{python}{ds.select}), or a boolean, where \mintinline{python}{False}
refers to no selection and \mintinline{python}{True} to the default selection.
This is useful for selections that are computationally expensive or selections
that are frequently used. In the current implementation, a named selection will
consume one byte per row, leading to a memory usage of 1~GB for a dataset
containing $10^9$~rows. Note that no copies of the data are being made, only a
boolean mask for the selection is created.

Often, a part of the data will not be used at all as part of preprocessing or
cleaning up. In this case we want a particular selection to be always applied,
without making a copy of the data. We refer to this as filtering, and is
similarly done as in \Pandas, e.g. \mintinline{python}{ds_clean = ds[ds.x_error
< 10]}). The filtering feature is implemented in exactly the same way as the
selections, except that a filter will always be applied, whereas a selection has
to be passed to a calculation explicitly each time.

A history of the expressions that define the selections is also kept, which
leads to less memory usage and enables users to undo and redo selections. In
addition, the expressions which define a selection can be stored on disk, making
the steps that led to a particular selection traceable and reproducible.

\subsubsection{Missing values}

It often happens that some samples in a column of a dataset lack an entry. Such
missing values are supported using \numpy's masked array type, where a boolean
is kept for every row of a column, specifying whether a value is missing or not.
For floating point numbers a NaN (Not a Number) values are also interpreted as
missing values.

\subsubsection{Units}
Optionally, a unit can be assigned to a column. Expressions based on columns
which all have units assigned to them will also result is an unit for that
expression. For visualization, the units can be used in the labelling of axis,
as is done in \texttt{vaex-viz} (see the x-axis labelling of the top panel of
Figure~\ref{fig:viz1d} for an example). The use of units fully relies on the
\texttt{Astropy} \Python package \citep{Astropy2013}.

\subsubsection{Statistics on N dimensional grids}
\label{sec:statistics}

One of the main features of \vaex is the calculations of statistics on regular
N-dimensional grids. The statistics, listed in Table~\ref{tab:algo} can be
computed for zero\footnote{A scalar, or single value.} or higher dimensional
grids, where for each dimension an expression, a bin-count and a range (a minimum
and a maximum) must be specified. All these calculations make use of the fast
N-dimensional binning algorithm which is explained in more detail in
Appendix~\ref{app:algo}. Each method can take one or more selections as
arguments, which will then stream over the data once, or as few times as
possible, giving optimal performance especially when the size of the data
exceeds that of the RAM.

\subsection{vaex-hdf5}

In order to achieve the performance estimated in Section~\ref{sec:main}, we need
to put some constraints on how the data is stored and accessed. If we use the
typical unbuffered POSIX read method, assuming all the data from disk is cached
in memory, we would still have the overhead of the memory copy, in addition to
the system call overhead. Alternatively, if the data is stored in a file format
that can be memory mapped, it will enable us to directly use the physical memory
of the operating system cache, eliminating unnecessary
overheads\footnote{Otherwise
we would be limited to half of the total memory bandwidth.}. Aside from the
memory mapping requirements, we also impose additional constrains on the
file format in which the data is to be stored. First, we require the data to be
stored in the native format of the CPU (IEEE 754), and preferably in the native
byte order (little endian for the x86 CPU family). Our second requirement is that the
data needs to be stored in a column based format, meaning that the datum of the next
row is in the next memory location. In cases where we only use a few columns,
such as for visualization, reading from column based storage is optimal since
the reading from disk is sequential, giving maximum read performance.

The well known and flexible file format \hdf has the capabilities to do both
column based storage and to store the data in little and in big endian formats.
The downside of the \hdf format is that it can store almost anything, and there
are no standards for storing meta information or where in the file a table
should be stored. To reconcile this issue, we adopted the VOTable as an example
\citep{std:VOTABLE}, and also implemented support for Unified Content
Descriptors \citet[][UCD]{std:UCD}, units and descriptions for every column,
and a description for the tables. Having UCDs and units as part of the column
description allows the software to recognize the meaning of the columns, and
suggest appropriate transformations for example. The layout of the file is
explained in more detail in \url{https://vaex.io}.

Although \vaex can read other formats, such as FITS, ascii or VOTable, these
require parsing the file or keeping a copy in memory, which is not ideal for
datasets larger than $\gtrsim 100$~MB. For superior performance, users can
convert these formats to \hdf using \vaex. An intermediate solution is the
column based FITS format, which we will discuss in section \ref{sec:vaex-astro}.

\subsection{vaex-viz} \label{sec:vaex-viz}

\begin{figure}
\begin{center}
\begin{tabular}{l}
\includegraphics[scale=0.45]{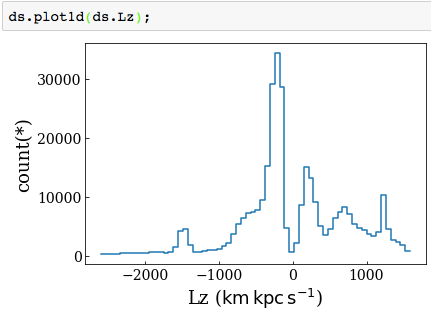} \\
\includegraphics[scale=0.45]{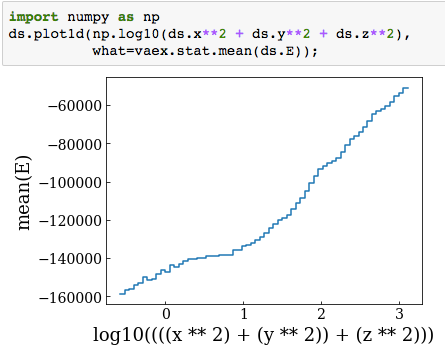} \\
\end{tabular}
\caption{Example of one dimensional visualization for the
\texttt{vaex-viz} package. \textbf{Top:} Histogram of \Lz, the angular
momentum in the $z$ direction. Because the units are specified in the data file
for this column, it will be included in the labelling of the x axis by default.
\textbf{Bottom:} Similar as above, but showing the mean of the energy \Energy
in each bin.}
\label{fig:viz1d}
\end{center}
\end{figure}

\begin{figure*}
\begin{center}
\begin{tabular}{ll}
\includegraphics[scale=0.45]{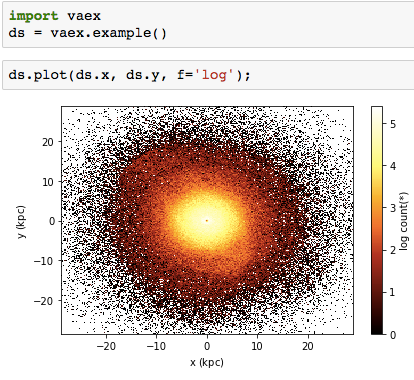} &
\includegraphics[scale=0.45]{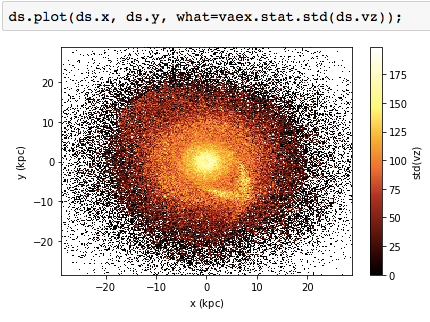} \\
\includegraphics[scale=0.45]{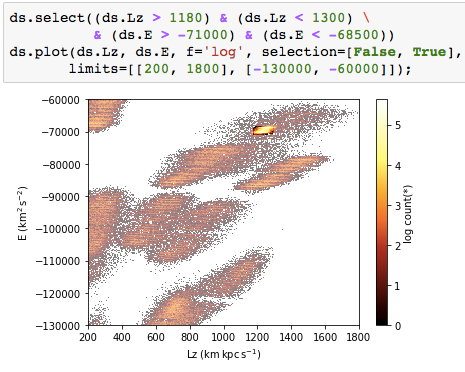} &
\includegraphics[scale=0.45]{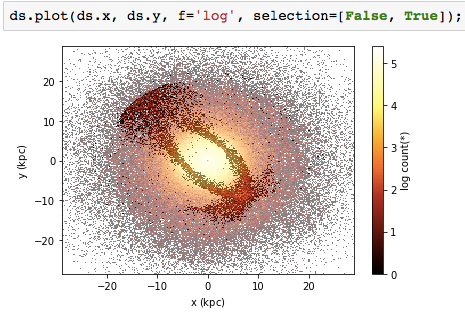} \\
\includegraphics[scale=0.45]{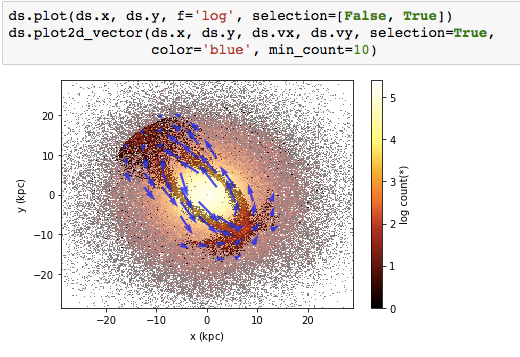} &
\includegraphics[scale=0.45]{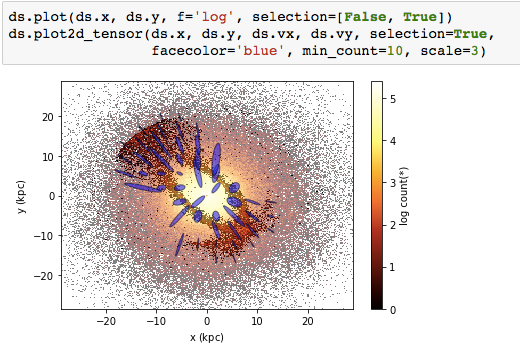} \\
\end{tabular}
\caption{Example of two dimensional visualization for the \texttt{vaex-viz}
package. \textbf{Top left:} two dimensional histogram, colour coded by a colour
map visualizing the counts in each bin. \textbf{Top right:} Similar to the left,
but showing the standard deviation in each bin. \textbf{Middle left:} A zoom-in
of the \Energy versus \Lz space, where a rectangular region is selected using
\vaex' selection mechanism. The underlying visualization of all the data is
slightly faded out to highlight the selection. \textbf{Middle right:} Similar to
the top left, but displaying the both the full dataset and the selection, where
the selected rectangular region corresponds to a satellite that is not fully
phase mixed. \textbf{Bottom left:} Similar as the middle right plot but
overlaying the mean velocity of the selection using vectors.
\textbf{Bottom right:} Similar as the bottom left, except now overlaying the
velocity dispersion tensor represented by ellipses.}
\label{fig:viz2d}
\end{center}
\end{figure*}

A core feature of \vaex is the visualization based on statistics calculated on
N-dimensional grids. The \texttt{vaex-viz} package provides visualization
utilising the \matplotlib library \citep{MPLHunter:2007}.

To showcase the main feature of \vaex, we use a random 10\% subset of the
dataset generated by \citet{Helmi2000MNRAS} which will be downloaded on the fly
when \mintinline{python}{ds = vaex.example()} is executed. This dataset is a
simulation of the disruption of 33 satellite galaxies in a Galactic potential.
The satellites are almost fully phase-mixed, making them hard to disentangle in
configuration space\footnote{The 3d positions.}, but they are still
separable in the space spanned by the integrals of motion: \Energy (Energy),
\Ltot (total angular momentum)\footnote{Although \Ltot is not strictly an
integral of motion in an axi-symmetric system, see \citet{Helmi2000MNRAS}.}, and
the angular momentum around the z axis \Lz. Even though this dataset contains
only $330\,000$~rows, it serves well to demonstrate what can be done with \vaex
while being reasonably small in size.

Larger datasets, such as 100\% of the \citet{Helmi2000MNRAS} dataset, the Gaia DR1 catalogue \citep{GaiaDR1cat}, or over 1 billion taxi trips in New York can be found at \url{https://vaex.io}.

For one dimensional visualizations, all statistics listed in
Table~\ref{tab:algo} can be plotted as a function of a single parameter or an
expression, where the count method will be used to create a histogram. An
example is shown in Figure~\ref{fig:viz1d}, where on the top panel we show a
regular one dimensional histogram for \Lz, while in the bottom panel we
visualize the mean energy \Energy in each bin in log radius. Note that the x-axis in the top
panel include a unit by default because these are include in the data file.

For two dimensional visualizations, we can display a two dimensional histogram
as an image. An example of this is shown in the top left panel of
Figure~\ref{fig:viz2d}, which shows a plot of $y$ versus $x$ (the positions of
the simulated particles) where the logarithm of the bin counts is colour-mapped.
Again, note the units that are included by default on the axes.
Similarly as for the one dimensional case, we can also display other statistics
in bins as shows in the top right panel of Figure~\ref{fig:viz2d}. Here instead
of showing the counts in bins, we display the standard deviation of the $z$
component of the velocity vector ($v_z$). This already shows some structure: a
stream that is not fully phase mixed can be readily seen.

In the middle left panel of Figure~\ref{fig:viz2d}, we create a selection
(a rectangle in \Energy and \Lz space), and visualize the full dataset and
the selection on top of each other. The default behaviour is to fade out the
full dataset a bit so that the last selection stands out. Since selections are
stored in the Dataset object, subsequent calculations and plots can refer to the
same selection. We do this in the right panel of the middle row of the same
Figure, where we show in a difference space ($y$ vs $x$) what the full dataset
and the selection look like. Here we can clearly see that the clump selected in
\Energy and \Lz space corresponds to a not fully phase-mixed stream, the same structure we
noticed in the top right panel in this Figure.

On the bottom left panel of Figure~\ref{fig:viz2d}, we use the same selection
but a different visualization. We first repeat the same visualization as in the
middle right panel, but overlay vectors, or a quiver plot. The
\mintinline{python}{plot2d_vector} method calculates a mean vector quantity on a
(coarser) grid, and displays them using vectors. This can give better insight
into vectorial quantities than for instance two separate density plots. The grid
on which the vectorial quantity is computed is much courser to not clutter the
visualization. Optionally, one can also colour-map the vectors in order to
display a third component, for instance the mean velocity in the $z$ direction.
Using a combination of the density map and the vector fields, one can plot up
to five different dimensions on the same Figure.

Similarly to a vectorial quantity, we can visualize a symmetric two dimensional
tensorial quantity by plotting ellipses. We demonstrate this on the bottom right
panel in Figure~\ref{fig:viz2d} using \mintinline{python}{plot2d_tensor}, where
we visualize the velocity dispersion tensor of the $x$ and $y$ components for
the same selection. In this visualization a diagonal orientation indicates a
correlation between the two velocity dispersion components and the size of the
ellipse corresponds to the magnitudes of the velocity dispersions.

\subsection{vaex-jupyter}

\begin{figure}
\begin{center}
\begin{tabular}{c}
\includegraphics[scale=0.4]{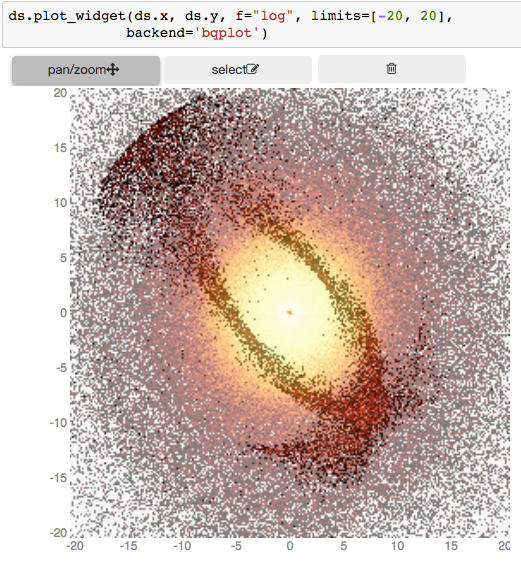}\\
\includegraphics[scale=0.4]{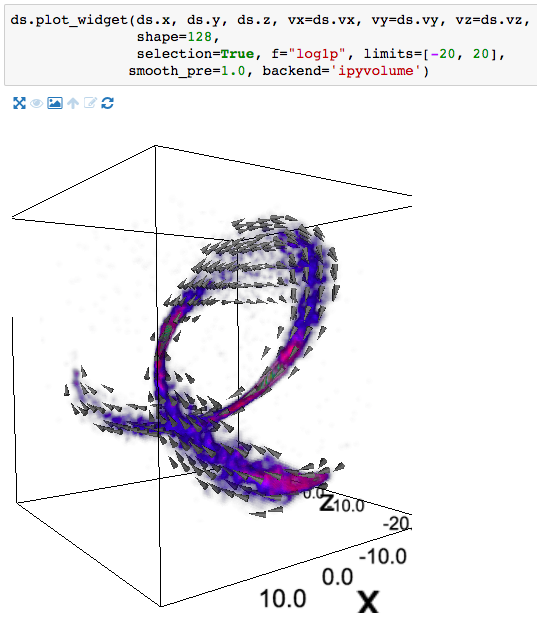}
\end{tabular}
\caption{\textbf{Top:} Screenshot of a Jupyter notebook using the widget
back-end to visualize the dataset interactively using panning, zooming and
selection with the mouse. The back-end used here is \texttt{bqplot}.
\textbf{Bottom:} A 3d visualization using the ipyvolume
backend, showing the stream discussed before, where the vector visualize the
mean velocity. \textit{An interactive version of this visualization can be found in the online version}.}
\label{fig:jupyter}
\end{center}
\end{figure}

The \vaex library, especially the visualization tools described in
Section~\ref{sec:vaex-viz} can also be used in combination with the Jupyter
(formerly IPython) notebook \citep{PER-GRA:2007}\footnote{Jupyter is the new
front end to the IPython kernel} or Jupyter lab, which is a notebook environment
in the web browser. A web browser offers more options for interactive
visualization and thus exploration of data compared to the static images which
are the default in \matplotlib. The Jupyter environment allows \vaex to work in
combination with a variety of interactive visualization libraries, mostly built
on \texttt{ipywidgets}\footnote{IPython widgets,
\url{https://ipywidgets.readthedocs.io/})}. For two dimensional visualization
\texttt{bqplot}\footnote{\url{https://github.com/bloomberg/bqplot}} allows for
interactive zooming, panning and on-plot selections as shown in the top panel of
Figure~\ref{fig:jupyter}.
\texttt{ipympl}\footnote{\url{https://github.com/matplotlib/jupyter-matplotlib}}
is an interactive back-end for \matplotlib, but unlike \texttt{bqplot}, the
visualization is rendered in the kernel as opposed to in the browser,
giving a small delay when rendering a plot. The
\texttt{ipyleaflet}\footnote{\url{https://github.com/ellisonbg/ipyleaflet}}
library can be used to overlay and interact with a geographical map.

For displaying the data in three dimensions we use
\texttt{ipyvolume}\footnote{\url{https://github.com/maartenbreddels/ipyvolume}},
which offers volume and isosurface rendering, and quiver plots using WebGL. A
big advantage of using WebGL in the Jupyter notebook is that it allows one to
connect to a remote server while running the visualization on the local computer,
a feature that is difficult to set up using OpenGL. The bottom panel of
Figure~\ref{fig:jupyter} shows an example of three dimensional volume
rendering including a quiver plot, created using the synergy between \vaex
and \texttt{ipyvolume} in the Jupyter environment. Note that in the three
dimensional version, especially the interactive version on-line, gives a much
clearer view on the orientation and direction of rotation of the stream
compared to the two dimensional version shown in Figure~\ref{fig:viz2d}.

\subsection{vaex-ui}
\label{sec:program}
\label{sec:vaex-ui}

\begin{figure*}
\begin{center}
\begin{tabular}{cc}
\includegraphics[scale=0.5]{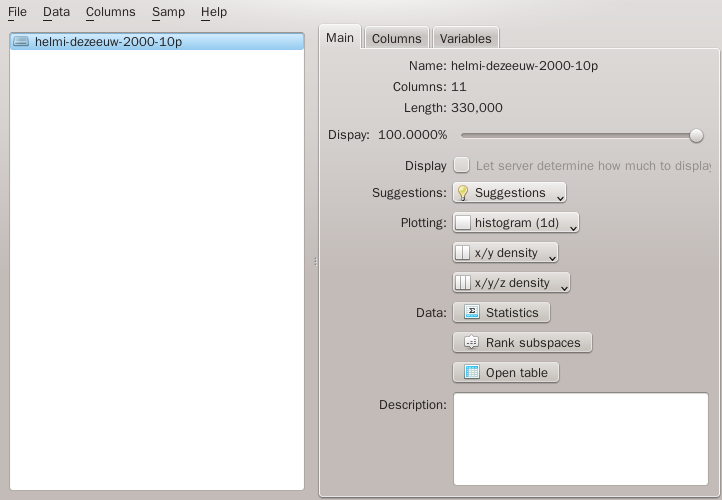} &
\includegraphics[scale=0.35]{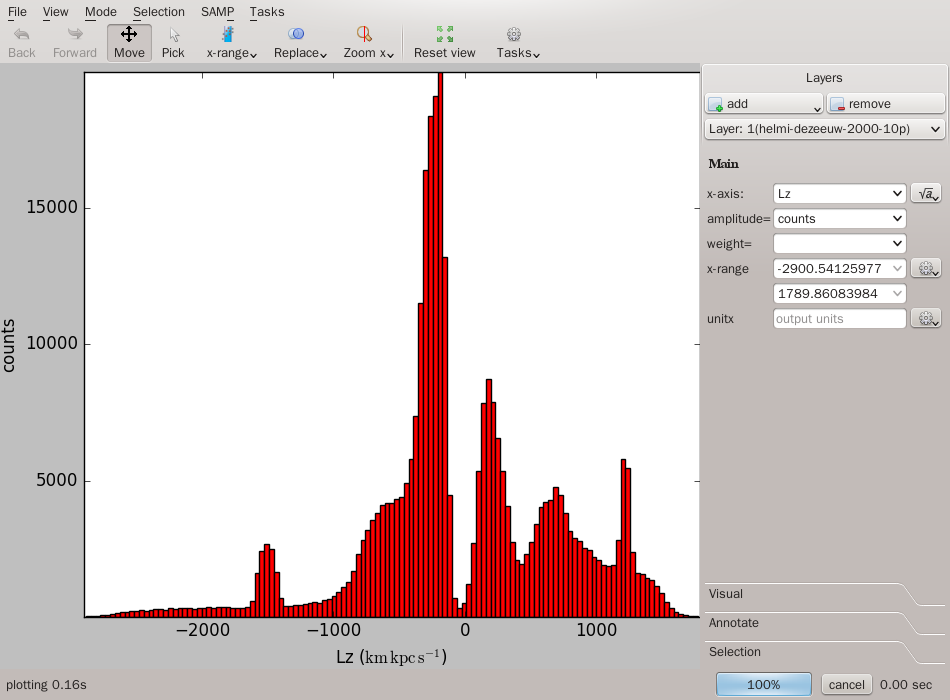} \\
\includegraphics[scale=0.35]{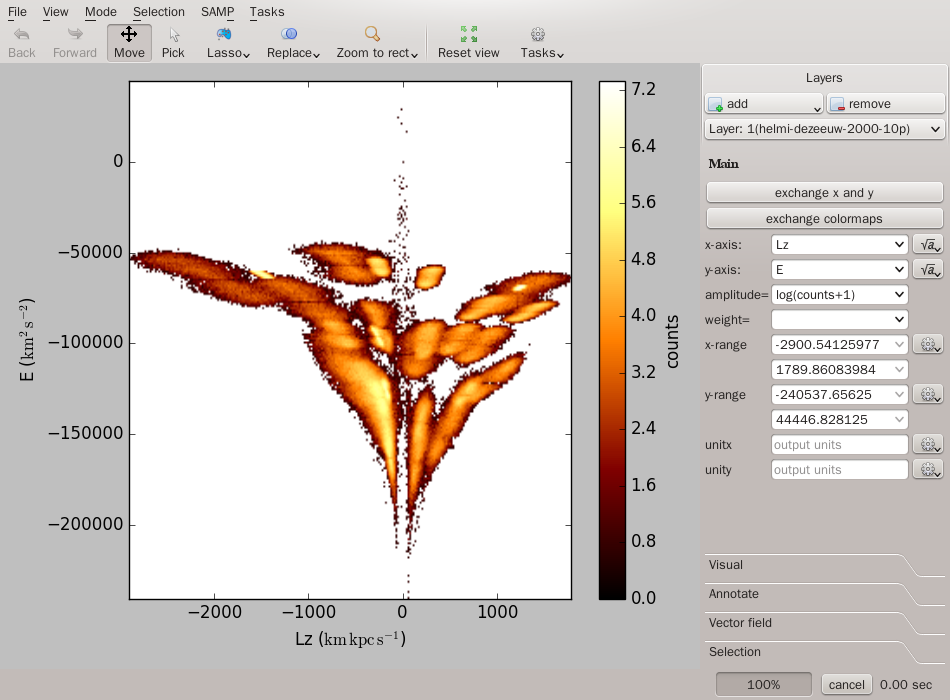} &
\includegraphics[scale=0.20]{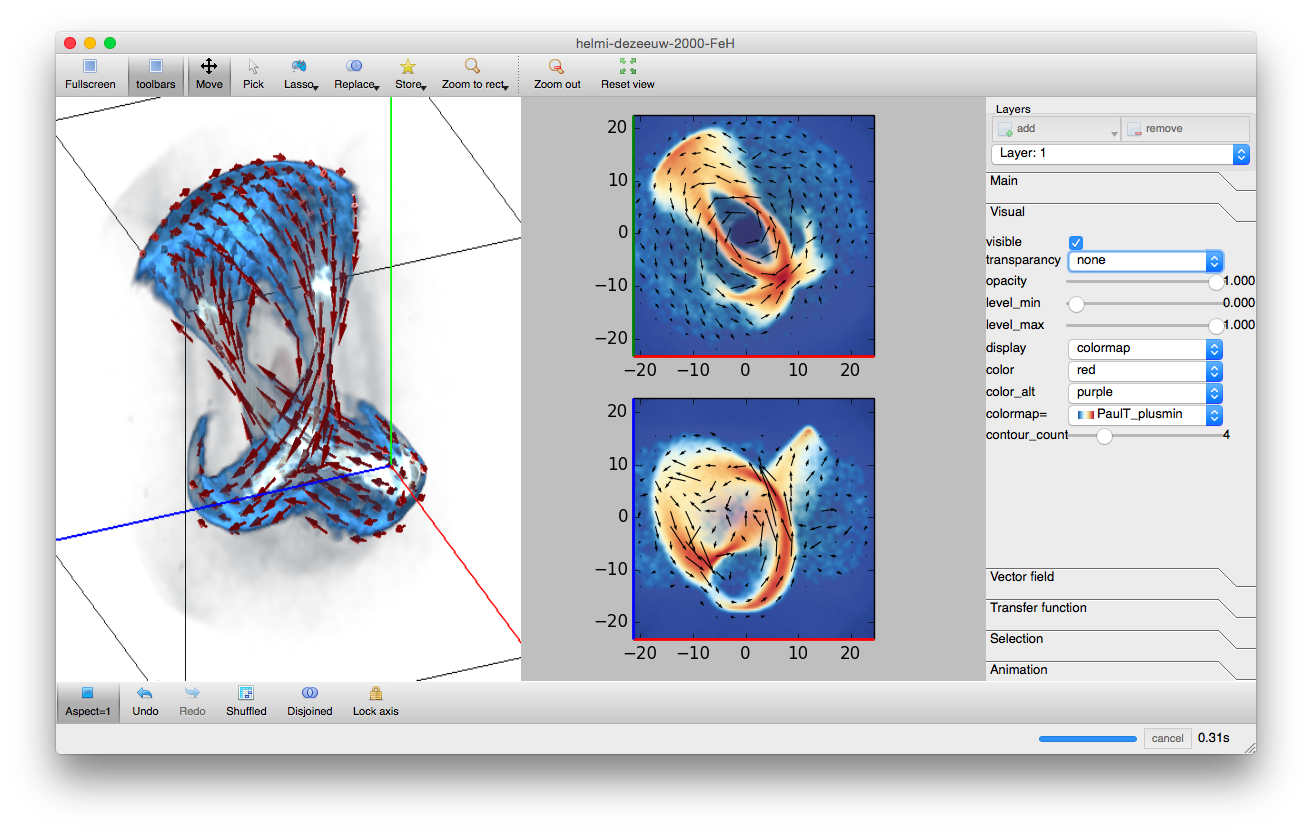}
\end{tabular}
\caption{\textbf{Top left:} A screenshot of main application window. On the
left it shows the open tables, and on the right the meta-data and operations that
can be performed on the table. \textbf{Top right:} One dimensional histogram
showing the \Lz distribution of our example dataset. \textbf{Bottom left:}
A two dimensional plotting window, showing \Energy versus \Lz.
\textbf{Bottom right:} A volume rendering of a stream in our example dataset in
Cartesian coordinates. The mean velocity field is also displayed with the help
of vectors.}
\label{fig:program}
\end{center}
\end{figure*}

\begin{figure*}
\begin{center}
\begin{tabular}{cc}
\includegraphics[scale=0.35]{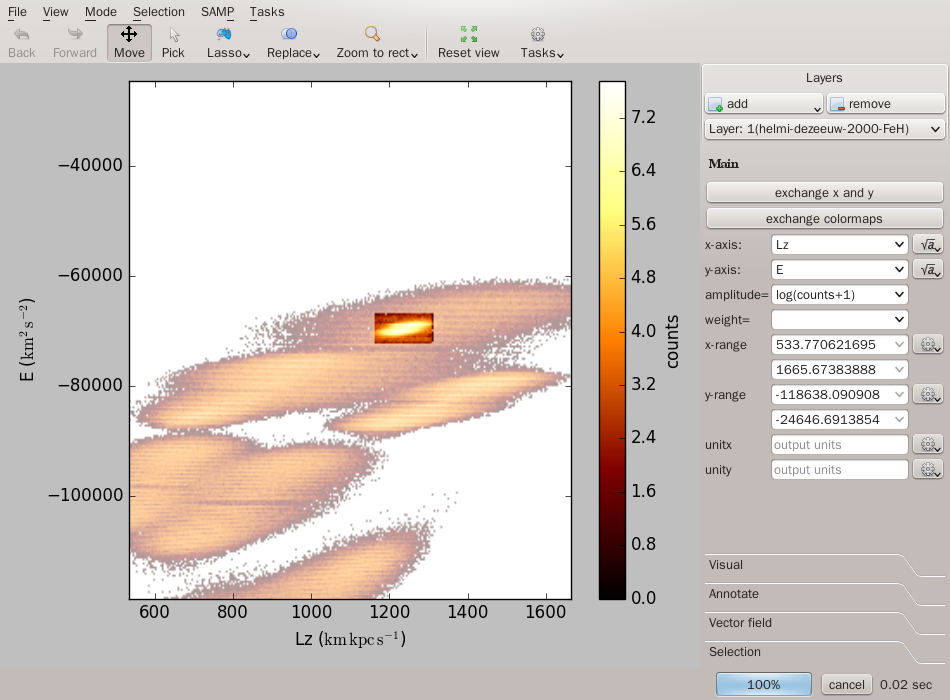} &
\includegraphics[scale=0.35]{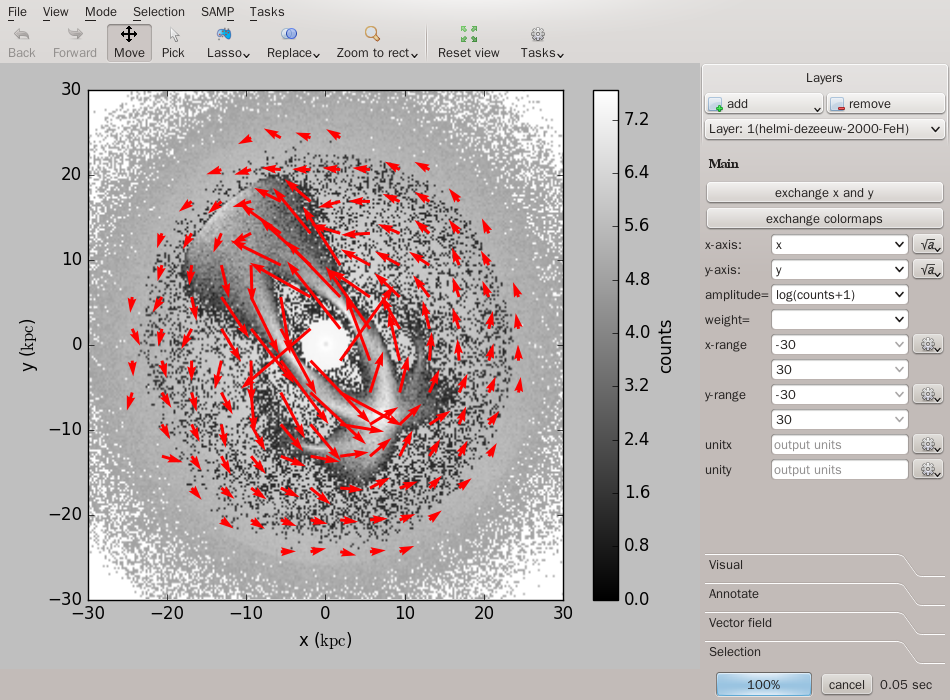}
\end{tabular}
\caption{\textbf{Left:} A two dimensional plot window, showing \Energy vs \Lz,
similar to the middle left panel of Figure~\ref{fig:viz2d}. \textbf{Right:} A two dimensional plot windows,
showing $x$ vs $y$, similar to the middle right panel of
Figure~\ref{fig:program}, but sharing the same selection as that shown in the
left panel, demonstrating the linked views feature.}
\label{fig:linked}
\end{center}
\end{figure*}

\begin{figure*}
\begin{center}
\begin{tabular}{cc}
\includegraphics[scale=0.45]{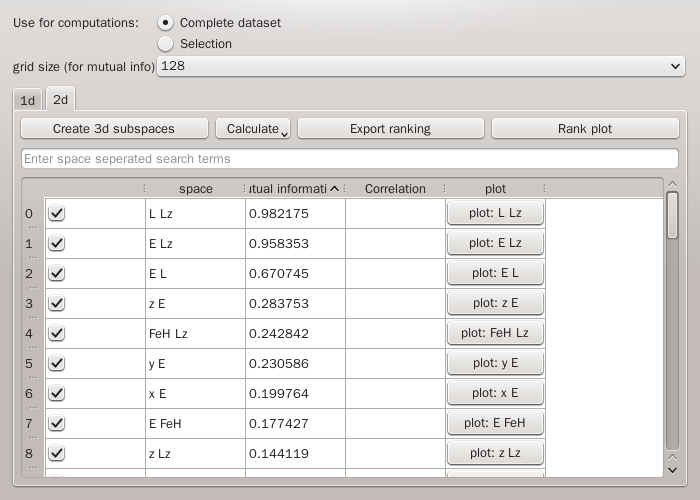} &
\includegraphics[scale=0.30]{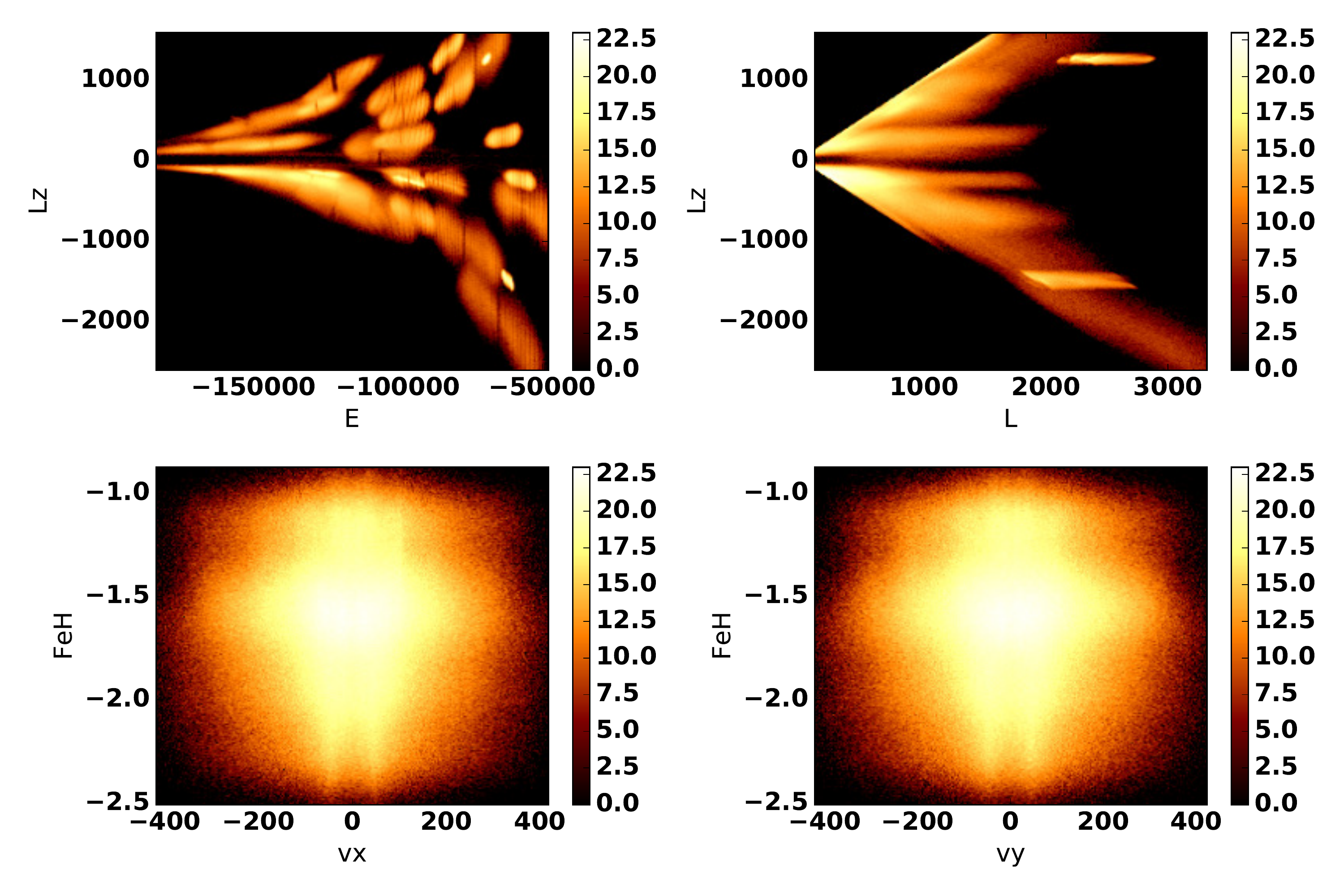}
\end{tabular}
\caption{\textbf{Left:} Ranking of subspaces my mutual information.
\textbf{Right:} The top row shows the two subspaces having the highest, while
the bottom panels show the two subspaces with the lowest mutual
information rank.}
\label{fig:ranking}
\end{center}
\end{figure*}

The \texttt{vaex-ui} package provides a graphical user interface that can be
used as a standalone program or used in the Jupyter environment. We focus now on
the standalone program. Upon starting, the program shows the open tables on the
left side, as shown in the top left panel on Figure~\ref{fig:program}. On the
right side it shows metadata information of the selected table, and it contains
buttons for opening windows that can visualize the dataset, give statistics on
the data, show the ranking of the subspaces, or display the whole table. The
next tabs shows the available columns, allows editing of units and UCDs, and
addition and editing of virtual columns. The third tab shows the variables that
can be used in expressions.

Similarly as in Section~\ref{sec:vaex-viz}, one can do one dimensional visualization
using histograms or statistics in regular bins. The top right panel
in Figure~\ref{fig:program} shows an example of this, where we plot the
histogram of \Lz for the example dataset as presented in
Section~\ref{sec:vaex-viz}. The plot shown on this Figure is interactive,
allowing zooming and panning with the mouse and keyboard, as well as interactive
selections. In the ``x-axis'' text box of this window, one can enter any valid
mathematical \Python expression, such as \mintinline{python}{log(sqrt(x**2+y**2+z**2))} for
example, where \mintinline{python}{x}, \mintinline{python}{y} and \mintinline{python}{z} are columns in the dataset we use. In addition to
the columns present in the dataset, one can also use any pre-defined virtual
columns or variables to compute user defined expressions. Apart from the
standard histograms that are based on the count statistic in
Table~\ref{tab:algo}, users can visualize other statistics, such as the mean or
standard deviation of an expression per bin.

For two dimensional visualizations, the program displays a two dimensional
histogram using an image, also similar to Section~\ref{sec:vaex-viz}. An example
of this is shown in the bottom left panel of Figure~\ref{fig:program}, which is
a plot of the \Energy versus the \Lz, where the logarithm of the bin counts is
colour-mapped. On this panel, one can see the individual satellites, each having
its own distinct energy and angular momentum. As in the case for the one
dimensional histogram, the entries for the x- and y-axis fields can be any valid
mathematical \Python expressions. Also similar to the one dimensional visualization, the statistics
listed in Table~\ref{tab:algo} can be visualized in the bins, now in two
dimensions. Vectorial and symmetric tensor quantities can be displayed as well,
in a similar manner to what is described in Section~\ref{sec:vaex-viz}.

The program also supports volume rendering using the OpenGL shading
language. We supports multi-volume rendering, meaning we can display both the
full dataset and a selection. In addition, one can over-plot vector fields in
three dimensions. The users have access to custom settings for the lighting and
for the transfer function. Navigation and selections are done in two dimensional
projection plots displayed alongside the panel that shows the three dimensional
rendering. An example of this visualization is shown in the bottom right panel
of Figure~\ref{fig:program}.

\subsubsection{Linked views}
The program also supports linked views \citep{Goodman2012}. This
feature makes all the active plots of the program linked to a single
selection, allowing for complex and interactive exploration of a dataset. To
demonstrate this concept, the left panel of Figure~\ref{fig:linked}, shows a
zoomed-in view of the bottom left panel of Figure~\ref{fig:program}, where we
have selected a particular cluster of stars in the subspaces spanned
by \Energy and \Lz. This is similar to what we have done in the middle left
panel of Figure~\ref{fig:viz2d}, except that we can now do the selection
interactively with the mouse. On the right panel of Figure~\ref{fig:program} we
see the how the selection looks like in configuration space, and we can readily
see that the selected clump in \Energy and \Lz space is in fact a stream. This
is confirmed the velocities of its constituents stars, which are displayed
with the help of a vector field overlaid on the same panel.

\subsubsection{Common features}

The windows that display the plots contain a number of options to aid users
in the exploration and visualization of their data. These include the setting of
an equal axis ratio, keeping a navigation history, undoing and redoing
selections, and numerous options to adjust the visualization settings. Users
also have the option to display multiple dataset on top of each other on the
same figure using the concept of layers, where where different blending
operations can be set by the user. There are also options for exporting figures
as raster (e.g. jpeg) or vector (e.g. pdf) graphics, as supported by the
\matplotlib library. It is also possible to export the binned data and with it a
script that reproduces the figure as seen on the screen. This enables users to
further customize their plots and make them publication ready.

\begin{figure*}
\begin{center}
\begin{tabular}{ccc}
\includegraphics[scale=0.55]{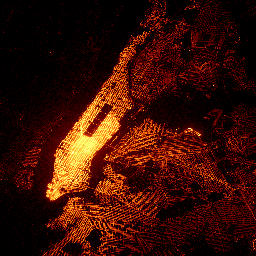}&
\includegraphics[scale=0.18]{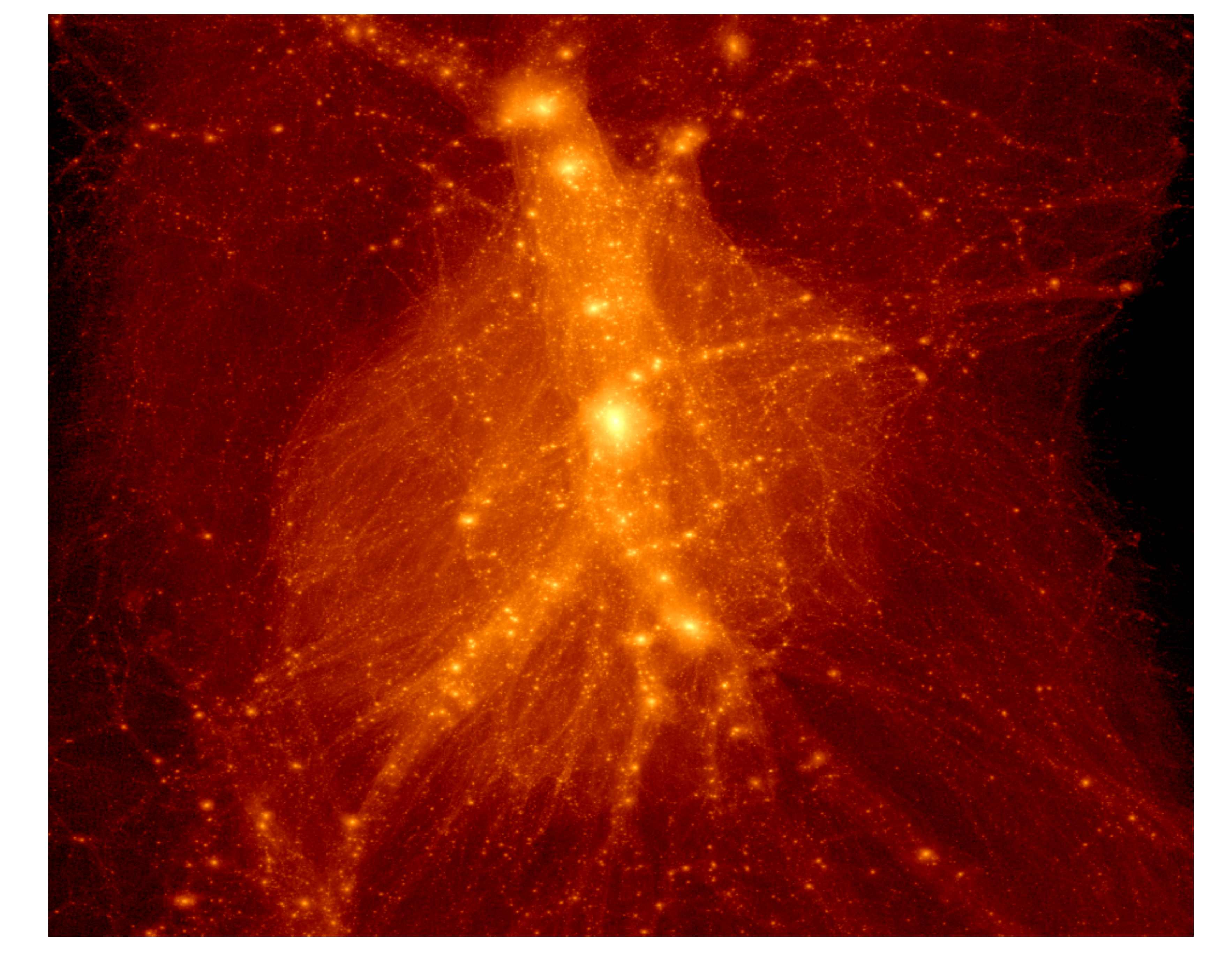} &
\includegraphics[scale=0.18]{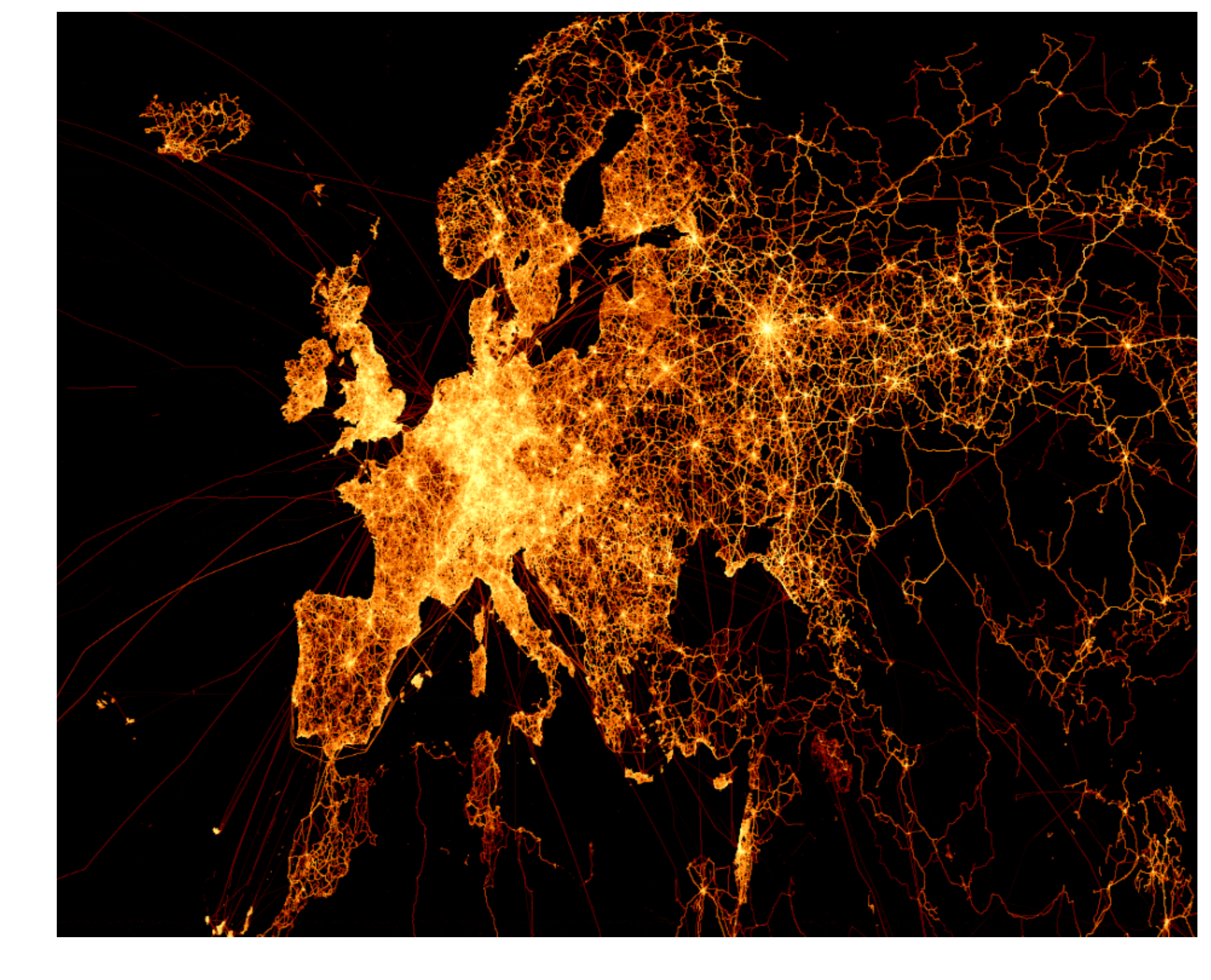}
\end{tabular}
\caption{Examples of density maps created from big datasets. \textbf{Left:}
Drop-off locations of the Yellow cab taxies in New York City between 2009 and
2015. In total $\sim 1$~billion GPS coordinates are plotted. \textbf{Middle:}
Density plot of particle position from the Aquarius pure dark matter
simulations \citep[][Aquarius A level 2]{Springel2008}, $\sim0.6$~billion
particles. \textbf{Right:} Open street map GPS data, showing $\sim 2$~billion
coordinates.}
\label{fig:other}
\end{center}
\end{figure*}

\subsubsection{Subspace exploration / ranking} \label{sec:mi}

Many tables nowadays contain a large number of columns. Thus, inspecting how
different quantities present in a dataset depend on each other, and finding
which subspaces, or combinations of those quantities in two or more dimensions
contain the most information via manual inspection by the user can be quite
tedious, and sometimes not feasible at all. To aid users in finding subspaces
that are rich in information, the program offers the option to rank subspaces
according to two metrics: the Pearson's correlation coefficient and the mutual
information. The calculation and interpretation of the Pearson's correlation
coefficient is well documented in classical statistics, and we will not discuss
it further.

The mutual information is a measure of the mutual dependence between two or more
random variables. It measures the amount of information one can obtain about
one random variable through the measurement or knowledge of another random
variable. In \vaex, the mutual information is obtained using the
Kullback-Leibler Divergence \citep[][KLD]{kullback1951}, and for two random
variables it is calculated via the expression:

\begin{equation}
I(X;Y) = \int_Y \int_X p(x,y) \log{\left(\frac{p(x,y)}{p(x)\,p(y)} \right) } \; dx \,dy
\label{eq:kld}
\end{equation}
where $p(x,y)$ is the joint probability distribution of the random variables
$x$ and $x$, while $p(x)$ and $p(y)$ are their marginalized probability
distributions. This quantity can be calculated for all, or a user defined subset
of subspaces having two or more dimensions. The mutual information effectively
measures the amount of information contained within a space spanned by two
or more quantities. On the left panel of Figure~\ref{fig:ranking} we show the
window listing all two dimensional subspaces in our example dataset, sorted by
mutual information. In the right panel of this Figure we show two subspaces that
have the highest and two subspaces that have the lowest rank according to their
mutual information, in the top and bottom rows respectively. One can readily see
that the spaces spanned by the integrals of motion (\Energy, \Ltot and \Lz) are
found to contains the most information, as expected.

\subsection{vaex-astro}
\label{sec:vaex-astro}
This subpackage contains functionality mostly useful for astronomical purposes.

\subsubsection{FITS}

Although FITS is not as flexible as \hdf, and is not designed to store data in
column based format, it is possible to do so by storing one row where each
column contains a large vector. \Topcat is using this strategy, and calls it
the \colfits format. However, the BINTABLE extension for \fits mandates that the
byte order to be big endian. This mean that the bytes need to be swapped before
use, which in \vaex gives a performance penalty of $\sim$30\%. For compatibility
with existing software such as \Topcat, we support the \colfits format both for
reading and writing.

\subsubsection{Common transformation}

The conversion between different coordinate systems is common in astronomy. The
\texttt{vaex-astro} package contains many of the most common coordinate
transformations, such as the conversion between positions and proper motions
between different spherical coordinate systems (Equatorial, Galactic, Ecliptic),
as well as the conversion of positions and velocities between spherical,
Cartesian and cylindrical coordinate systems. The transformations also include
the full error propagation of the quantities in question. If the dataset contains
metadata describing the units of the quantities, these will be automatically
transformed as needed, minimizing possible sources of error.

\subsubsection{SAMP}

\Vaex offers interoperability with other programs via the Simple Application
Messaging Protocol \citep[SAMP][]{SAMP2009ivoa.spec.0421B}. It can communicate
with a SAMP hub, for instance by running \Topcat's build in SAMP hub, which
can then broadcast a selection or objects that are `picked' to other programs.
\Vaex understand the `table.load.votable` message, meaning other programs can
send tabular data in the VOTable format to it. Although this transfer mechanism
is slow, it means that any data that can be read into \Topcat can be passed on
to \vaex. For example, one can download a VOTable from VizieR using the TAP
protocol \citep{std:TAP}, which is fully implemented in \Topcat, and than use
SAMP to transfer it to \vaex. The same is possible with any other program,
application or web service that supports SAMP.

\subsection{vaex-server}
\label{sec:vaex-server}

A dataset that consists of two columns with 1~billion rows filled with double
precision floating point values amounts to 16~GB of data. On the other hand, if
this dataset is binned on a $256\times256$ grid, which also uses double
precision floating points for its bin valuesm has the size of only 0.5~MiB.
Using a suitable compression this can be reduced further by a factor of
$\approx 10$. This makes it possible and practical to do calculations on a
server and transfer only the final results to a client.

A working client/server model is implemented in the \texttt{vaex-server}
package, making it possible to work with both local and remote datasets at the
same time. Also the program provided by \vaexui allows a users to connect to a
remote server. The server itself is completely stateless, meaning it does not
keep into memory the state of a remote client. This means that when a user
requests a statistic for a subsets of the data defined by a selection, the
client needs to send that selection to the server, and the server will compute
the selection at each request. The server can cache selections, speeding up
subsequent calculations of statistics for which the same subset of the data is
required. The benefit of having it stateless is that the server can be less
complex, and can be scaled vertically, meaning that more servers can be added
with a load balancer in front to scale up the workload.

Furthermore, clients have the option to let the server determine how much of
the data it will use, assuming the data is shuffled, to give a approximate
answer. The server will then estimate how much of the data should be processed
to return a result in a predefined amount of time, which is 1~second by default.
Clients that want to use a specific amount of data, up to 100\%, may need to
wait longer. Using this mechanism, the \texttt{vaex-server} can handle up to 100
requests per second on a single computer.

\subsection{vaex-distributed}

Many of the statistics that \vaex can compute, such as the mean or variance
can be written as a linear combination of smaller chunks
of the data used. Thus it is possible to distribute the computations to
different computers, each performing a part of the work on a smaller data chunk,
and combine the results at the end. The \texttt{vaex-distributed} package makes
this possible. With its use, we manage to perform computations 10~billion
$(10^{10})$ rows per second on a cluster of 16 low-end computers. This
demonstrates that \vaex can be scaled up to even larger datasets of the order of
$\approx 10-100$~billion rows with the help of a computer cluster, even if
such a cluster is not composed of modern computers. Note that not all
functionality is supported, as this is only a proof of concept.

\subsection{vaex-ml}

Build on top of vaex, is another proof of concept package called \texttt{vaex-ml}, which combines machine learning with \vaex' efficient data handling. Comparing \texttt{\vaex-ml} a k-means clustering algorithm to that of sklearn \citep{sklearn}, we are about $5 \times$ faster and have low memory impact since \vaex does not need to load the dataset in memory nor copies it. In both cases all cores are used. For PCA we are almost $7 \times$ faster, but \vaex by default uses multi-threading for the calculation of the covariance matrix. Note that the sklearn implementations of both PCA and k-means are limited to ~10~million rows in order to avoid using the swapdisk, while \vaex happily works through the $100$ million rows. We furthermore integrated \vaex with \texttt{xgboost} \citep{xgboost} to make boosted tree models easily available. 
Note that the \texttt{vaex-ml} source code is available and is free for personal and academic usage.

\subsection{Any large tabular dataset}

We would like to emphasize that, even though the main motivation for creating
\vaex was to visualize and explore the large \emph{Gaia} catalogue, \vaex is an
ideal tool to use when working with any other large tabular dataset. To
illustrate this point, in Figure~\ref{fig:other} we visualize three large
datasets. The leftmost panel is a density plot of $\sim 1$~billion drop-off
locations made by the Yellow cab taxies in New York City for the period between
2009 and 2015. The middle panel on the same Figure shows the positions for
0.6~billion particles from the pure dark matter Aquarius simulation
\citep[Aq-A, level2][]{Springel2008}. The right panel in Figure~\ref{fig:other}
displays the Open street map GPS data over Europe, made by $\sim 2$~billion GPS
coordinates. These plots demonstrate that \vaex is a suitable tool for
exploration and visualization of any large tabular datasets regardless whether
they are related to astronomy or not.

\section{Conclusions}
\label{sec:conclusions}

In the future datasets will grow ever larger, making the use of statistics
computed on N-dimensional grids for visualization, exploration and analysis more
practical and therefore more common. In this paper we introduced \vaex, a tool
that handles large datasets, and processes $\sim 1$~billion rows per second on a
single computer. The \vaex \Python library has a similar API to that of \Pandas,
making for a shallow learning curve for new users, and a familiar, easy to
understand style for more experiences users. Built on top of many of the \vaex
packages is \vaexui, which provides a standalone program allowing data
visualization in one, two and three dimensions, interactive exploration of the
data such as panning and zooming, and visual selections. By combining the \vaex
program, which can be used for a quick look at the data, with the \vaex library
for more advanced data mining using custom computations, users can be quite
flexible in the manner in which they explore and make sense of their large
datasets.

In the era of big data, downloading a large dataset to a local machine may not
always be the most efficient solution. Using \texttt{vaex-server}, a dataset can
be worked on remotely, only sending the final products, such as the (binned)
statistics to the user. Combining multiple server in \texttt{vaex-distributed}
allows \vaex to scale effortless to $\sim 10$~billion rows per second on small
cluster of a dozen computers.

We have demonstrated many of the features of \vaex using the example dataset
from \citep{Helmi2000MNRAS}. Visualization of statistics in one, two and three
dimensions, using the full and subsets (selections in \vaex) of the data as well
as the visualization of vectorial and (symmetric) tensorial quantities. All of
these calculations and visualizations will scale to datasets with billions of
rows making \vaex the perfect tool for the Visualization And EXploration
(vaex) of the \emph{Gaia} catalogue \citep{GaiaDR1cat}, and even more so for the
upcoming data releases or future missions such as LSST. The first data release
of \emph{Gaia} is available in \hdf format at \url{http://vaex.io}, and we plan
to do so as well for the second data release.

\Vaex is open source, and available under the MIT license.
Contributions are welcome by means of pull requests or issue reports on
\url{https://github.com/maartenbreddels/vaex}. The main website for \vaex is
\url{https://vaex.io}.

\section*{Acknowledgments}
MB and JV thank Amina Helmi for making this work possible. MB thanks Yonathan Alexander
for pushing me to create a more user friendly API. MB and JV AH are
grateful to NOVA for financial support. 
This work has made use of data from the European Space Agency (ESA) mission
\emph{Gaia} (\url{http://www.cosmos.esa.int/gaia}), processed by the \emph{Gaia}
Data Processing and Analysis Consortium
(DPAC, \url{http://www.cosmos.esa.int/web/gaia/dpac/consortium}). Funding for
the DPAC has been provided by national institutions, in particular the
institutions participating in the \emph{Gaia} Multilateral Agreement.

\bibliography{vaexpaper}

\appendix

\section{Binning algorihm}
\label{app:algo}

\begin{figure*}
\begin{minted}{python}
# This is equivalent code for the c code, but written in Python for readability
# It is for 2d only, 0, 1, and >= 3 dimensional are a generalization of this
# but more difficult to read
import numpy
def operation_count(input, aux):
    if aux is not None and numpy.isinf(aux):
        return input
    else:
        return input+1

def operation_minmax(input, aux):
    if  numpy.isinf(aux):
        return input
    else:
        return min(input[0], aux), max(input[1], aux)

def operation_moment_012(input, aux):
    if  numpy.isinf(aux):
        return input
    else:
        return [input[0] + 1, input[1] + aux, input[2] + aux**2]

def statistic2d(grid, x, y, aux, xmin, xmax, ymin, ymax, operation):
    grid_width, grid_height = grid.shape[:2] # get dimensions of the 2d grid
    for i in range(len(x)):  # iterator over all rows
        # normalize the x and y coordinate
        norm_x = (x[i] - xmin) / (xmax-xmin)
        norm_y = (y[i] - ymin) / (ymax-ymin)
        # check if the point lies in the grid
        if ( (norm_x >= 0) & (norm_x < 1) &  (norm_y >= 0) & (norm_y < 1) ):
            # calculate the indices in the 2d grid
            index_x = numpy.int(norm_x * grid_width);
            index_y = numpy.int(norm_y * grid_height);
            # apply the operation
            grid[index_x, index_y] = operation(grid[index_x, index_y],
			      aux[i] if aux is not None else None)

# To make a 2d histogram of 10 by 20 cells:
# data_x and data_y are 1d numpy arrays containing the data, and
# xmin, xmax, ymin, ymax define the border of the grid
shape = (10,20)
counts = np.zeros(shape)
statistic2d(counts, data_x, data_y, None, xmin, xmax, ymin, ymax, operation_count)

# To get a 2d grid with a min and max value of data_x at each cell
minmax = np.zeros(shape + (2,))
minmax[...,0] = np.inf
# Infinity and -infinity are good initial values since they will always be bigger
# (or smaller) than any finite value.
minmax[...,1] = -np.inf
statistic2d(minmax, data_x, data_y, data_x, xmin, xmax, ymin, ymax, operation_minmax)

# calculate the standard deviation on a 2d grid for x by calculating the count, the
# sum of x and the sum of x**2 at each cell
moments012 = np.zeros(shape + (3,))
statistic2d(moments012, data_x, data_y, data_x, xmin, xmax, ymin, ymax, operation_moment_012)
# then calculate the raw moments
moments2 = moments012[...,2] / moments012[...,0]
moments1 = moments012[...,1] / moments012[...,0]
# and finally the standard deviation (non stable algorihm)
std = numpy.sqrt((moments2 - moments1**2))


\end{minted}
\caption{Python code equivalent of our generalized Nd binning code, but only for 2d for clarity. \label{code:algo}}
\end{figure*}

The binning algorithm in \vaex is a generalization of a one dimensional binning
algorithm to N dimensions. It also supports custom operations per bin, on top
of simply counting the number of samples that a bin contains. The algorithm
itself is written in \Clang and \Cpplang. For presentation purposes, we rewrote
it in pure \Python and how it in Figure~\ref{code:algo} below. We consider
\Python code to be equivalent to pseudo code and thus self explanatory. The
example includes the calculation of the counts, the minimum and maximum
statistics, as well as the standard deviation on a regular two dimensional grid.

\end{document}